\documentstyle[aps,prl,multicol,epsf]{revtex}
\def\beq{\begin{equation}}
\def\eeq{\end{equation}}
\begin{document}                
\title{Non local screening in a vortex line liquid}
\author{M. V. Feigelman{$^1$} and L. B. Ioffe$^{2,1}$ }
\address{{$^1$}Landau Institute for Theoretical Physics, Moscow, Russia}
\address{{$^2$}Physics Department, Rutgers University, Piscataway, NJ 08855}
\maketitle
\begin{abstract}
We show that the recent experiments \cite{safar} reporting the onset of the
non-local conductivity in the vortex state of $YBaCuO$ single crystals
indicate the presence of a new liquid phase of vortices.
This phase is intermediate between the normal metal and the Abrikosov lattice.
We use the mapping of the vortex problem to the problem of bose liquid to
determine theoretically the properties of the proposed vortex liquid phase and
compare them with the data.
\end{abstract}
%
\begin{multicols}{2}

In a recent Letter H. Safar {\em et al} reported \cite{safar} the results of
the transport measurements in the mixed state of $YBaCuO$ single crystals with
the field applied along $c$-direction of the crystal performed in a
transformer geometry shown in Fig 1.
Two sets of measurements were done.
In the first one the current was injected through the pair of contacts $(1,4)$
and the potential was measured between contacts $(2,3)$ or $(6,7)$.
In the second set the current was injected through contacts $(1,5)$ and the
potential was measured between contacts $(2,6)$, $(3,7)$ or $(4,8)$.
In the first setup the current flows predominantly along $ab$ plane of the
crystal whereas in the second setup the current flows mostly in $c$ direction.

\begin{figure}
\centerline{\epsfxsize=5cm \epsfbox{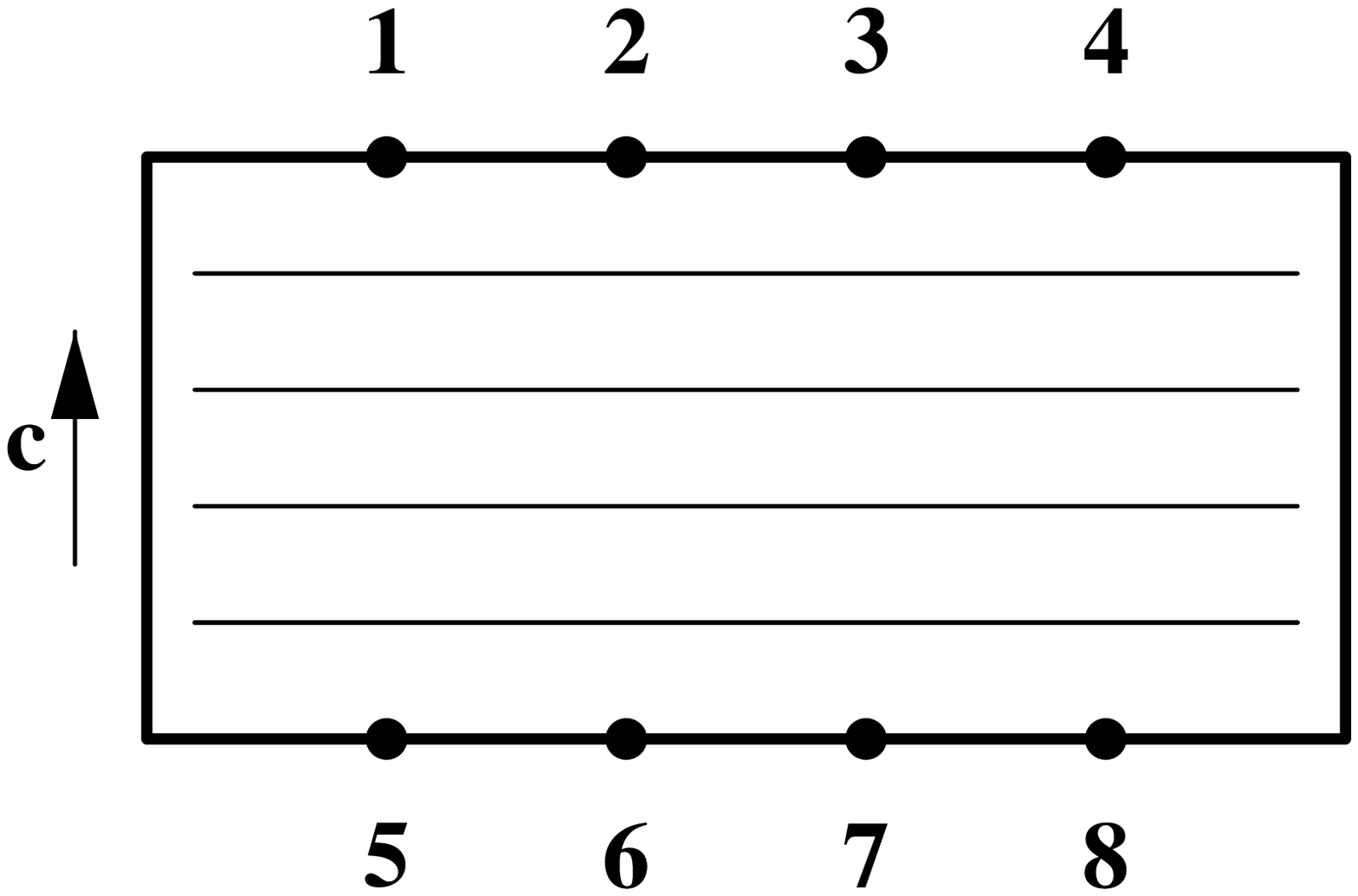}}
\end{figure}
Fig. 1. Geometry of dc transformer experiment. Heavy dots denote contacts.

\vspace{0.2in}

It was found \cite{safar} that in the first setup the potential drop between
contacts $(6,7)$ is smaller than the drop between $(2,3)$ at high temperatures,
but becomes undistinguishable from it below some critical temperature $T_{th}$.
If interpreted in terms of the apparent resistivity ratio  $\rho_c/\rho_{ab}$
such observation implies that this ratio tends to zero at
$T\rightarrow T_{th}(H)$.
This temperature $T_{th}(H)$ is significantly higher than the transition
temperature  $T_g(H)$ defined by $\rho_{ab}(T_g)=0$.
However, the experiments of the second setup showed that the ratio of the
potentials $V_{48}/V_{26}$ becomes much {\em larger}  when $T\rightarrow
T_{th}(H)$ indicating that apparent ratio $\rho_c/\rho_{ab}$ extracted from
these measurements tends to infinity.
Finally, an independent set of measurements indicate that resistivity in
$c$-direction extracted from measurements with the uniform current
$\rho_c(T_{th})=0$ \cite{bariloche}.

Safar {\em at al} interpreted their results as an evidence for non local
transport in the mixed state; in this note we shall show that this non local
transport may be a signature of a phase transition into the novel vortex
liquid state characterized by a non-zero phase rigidity in the
direction of the magnetic field \cite{theory0,theory}.
We shall argue that $T_{th}$ should be identified with the transition
temperature into the intermediate vortex liquid state.

We begin with the estimates of the parameters of the vortex system in the
regime of the experiments \cite{safar,bariloche}.
We use the value of the penetration length $\lambda=1400 \AA$ and anisotropy
ratio $M/m=50$ to estimate a vortex entanglement length (cf.\cite{nelson})
in $c$-direction:
$L = 1.5 (1-t)/B \; \mu m $ where $t=T/T_c$ and $B$ is measured in Tesla which
gives $L \sim 0.03 \; \mu m$ in the conditions of \cite{safar,bariloche}.
Thus, the vortices are entangled on a much shorter scales than the sample
thickness ($0.03 \;mm$).
This estimate rules out the simple explanation of the experiment \cite{safar}
that each individual vortex remains straight on the scales of the sample
thickness.
The absence of current dissipation in $c$-direction shows that although
vortices wander at short scales they remain relatively straight at large
scales due to collective effects.
The dissipation of the uniform in-plane current shows that the vortex
lattice is not formed.

In order to describe the liquid of vortices state quantitatively we mapped the
vortex problem  to the model of strongly interacting two dimensional bosons
\cite{theory}.
We shown that the problem of Gibbs equilibrium state of the system of
infinitely long vortices at temperature $T$ is formally equivalent to the
quantum ground state problem of two-dimensional interacting Bose liquid
(with the "2D Plank constant" $\hbar_{2D} = T$) at zero temperature.
The direction of the vortex space along the magnetic field
(${\hat z} \parallel {\bf c}$) plays the role of imaginary time of  the 2D
quantum theory.
The interaction between bosons in the quantum model has two parts:
static 2D Coulomb part and dynamic transverse current-current
interaction mediated by the fluctuating 2D "magnetic" field (cf.
\cite{theory0,theory}).
This mapping identifies the Abrikosov lattice and  the crystal of bosons, the
normal metal and the Bose superfluid liquid.
In \cite{theory} we also found a new vortex liquid phase between Abrikosov
lattice and normal metal which is mapped into the "normal" (i.e. nonsuperfluid)
liquid of bosons.

Here we shall use this mapping to find the response of the vortex phase in the
conditions of experiment \cite{safar}.
In \cite{theory0,theory} we derived the duality relations  between the response
functions in the vortex system and the response function for the boson system
and found that the new intermediate phase has non-zero superfluid density in
$c$-direction: $j_c = - n_s^{cc}A_c$ but has a finite resistivity to a {\em
uniform} current in $ab$-directions.
The phase transition line normal metal-new phase predicted in Ref.
\cite{theory}, Fig. 1a is surprisingly close to $T_{th}(H)$ line reported in
\cite{safar,bariloche}.
Below we show that in this state the {\em non-uniform} in-plane current
which satisfies $\int j_{ab} dz = 0$ is non-dissipative.
This resolves the apparent contradiction of observations because in the
set up yielding $\rho_{ab}/\rho_c \rightarrow 0$ the in-plane current obeys
$\int j_{ab} dz = 0$.

We  prove the existence of a non-dissipative current $j_{ab}(q_z\neq 0)$
repeating the derivation of response functions (cf. Section VI.A.1 of
\cite{theory})  for the in-plane currents.
 \beq
\begin{array}{rr}
D(q) = \langle A_{a}(q)A_{a}(-q) \rangle = \frac{4\pi T}{q^2 +
{\cal P}_{\parallel}(q)} ;\\
{\cal P}_{\parallel}(q) = \frac{1}{\lambda^2} \frac{q_z^2}{q^2 + g^2
\Pi_{\parallel}(q)}
\end{array}
\label{D}
\eeq
where $\Pi_{\parallel}(q)$ is the longitudinal response of the dual bose
liquid and $g^2=\frac{\phi_0^2}{4\pi\lambda^2}$ is its interaction constant
i.e. the analog of the electric charge $e_{2D}$.
Here and below we explicitly consider only isotropic superconductors; we shall
restore anisotropy factor $m/M$ only at the very end using general scaling
arguments \cite{scaling}.

The intermediate phase corresponds to the normal liquid of bosons.
Since $q_z$ play the role of the frequency in the bose model
$\Pi_{\parallel}(q)$ is strongly dependent on the ratio $q_z/q_{\perp}$.
In the limit $q_\perp \rightarrow 0$  the longitudinal response of the boson
system is described by the finite effective 'conductivity' $\sigma$:
$ g^2 \Pi_{\parallel} (q_z) = \sigma |q_z|$.
To estimate the effective conductivity we note that normal bose liquid is
realized in the regime of the strong interaction when the dimensionless
interaction parameters are the order of unity \cite{theory}.
In this regime the only combination of parameters with dimensionality of
conductivity is the "quantum" conductivity $\sigma_Q = e_{2D}^2/h_{2D}$.
Identifying $g \rightarrow e_{2D}$, $T \rightarrow \hbar_{2D}$, we
estimate  $\sigma \sim \frac{g^2}{2\pi T}$.
Using this expression for $\Pi_{\parallel} (q_z)$ we get the
correlator of the in-plane electromagnetic vector potential:
\beq
D(q_z) = \frac{4 \pi T }{q_z^2 + \Lambda^{-1} |q_z|}
\label{D 2}
\eeq
where $\Lambda = \lambda^2 \sigma = \frac{\Phi_0^2}{8\pi^2 T}$.

Eq. (\ref{D 2}) shows that the in-plane magnetic field decays at large
distances ($z \gg \Lambda$) as $B(z) \sim B_0
\frac{\Lambda}{z} $ in this state, being screened on a scale of $\Lambda$ by a
non-dissipative current $j \sim \left(\frac{\Lambda}{z}\right)^2$ along the
boundary.
To see this, consider the effect of the external current flowing in the $x$
direction on the edge of the sample $j_x^{ext}(z) = J^{ext} \delta(z)$.
According to Eq.(\ref{D 2}) it  induces magnetic field
$B_y(q_z) = q_z\cdot D(q_z) \cdot J^{ext} \propto 1/(q_z + sgn(q_z)
\Lambda^{-1})$.
After integration over $q_z$ one gets the induced field $B_y(z) \sim B_0
\frac{\Lambda}{z}$ and the shielding current $j_x(z) \propto dB_y/dz$.
Note that in the superfluid phase of bose-liquid one would get
$\sigma \propto 1/q_z$ and $\Pi_{\parallel}(q_z) \sim const$, so the
electromagnetic correlation function would acquire its usual for
normal metal form $D(q_z) \propto q_z^{-2}$.
The universal length scale $\Lambda$ is about $400 \; \mu m$ at $T=90 \; K$
which is larger than the sample thickness $d$
( note that the rescaling \cite{scaling} taking into account anisotropy does
not affect the relation between $\Lambda$ and $d$, since both quantities scale
in the same way).
The existence of non-dissipative in-plane current at $q_z \neq 0$ in the
vortex liquid implies
that $\rho_{xx}(q_z \neq 0) = 0$ at $T < T_{th}(H)$.
We interpret observations \cite{safar,bariloche} as a signature of  vanishing
of $\rho_{xx}(q_z \neq 0)$ at $T \rightarrow T_{th}(H)^+$,
which supports a phenomenological description proposed in Ref. \cite{huse}.

The Eq.(\ref{D 2}) for $D(q_z)$ was derived in the limit of in-plane
homogeneous current, i.e. $q_{\perp} \rightarrow 0$.
Using the conventional diffusion form for the density-density correlator
\[
\Pi_{\parallel}({\bf q}) \sim \frac{\sigma q_z^2}{|q_z| + {\cal
D}q_{\perp}^2}
\]
we see that Eq.(\ref{D 2}) remains valid  if $q_{\perp}^2 \ll q_z / {\cal D}$.
Here ${\cal D}$ has the meaning of ``diffusion coefficient'' associated with
``conductivity'' $\sigma$.
Thus,
$\sigma \approx e^2_{2D}{\cal D} \left(\frac{m_{2D}}{2\pi\hbar_{2D}^2}\right)$
where the last factor in parenthesis is  the density of states for 2D
particles with mass $m_{2D} = (\Phi_0/4\pi\lambda)^2$.
The ``quantum-limit'' expression for $\sigma$, which we used above means that
the mean free path of bosons is of the order of their separation, i. e.
${\cal D} \approx \hbar_{2D}/m_{2D}$.

Now we estimate the relative size of wavevectors $q_z$ and $q_\perp^2$ for the
conditions of the  experiment \cite{safar}.
After the rescaling \cite{scaling} takes into account with the mass anisotropy
factor $m/M = 0.02$, the relevant wavevectors become
$q_z \approx \sqrt{m/M}/d$, $q_{\perp} \approx 1/d_{plane}$, with the
sample thickness $d = (3-6)\cdot 10^{-3} cm$ and the relevant  lateral
dimension $ d_{plane} \approx 0.05 cm $.
With the above estimates for ${\cal D}$, the condition $q_{\perp}^2 \ll q_z /
{\cal D}$ is satisfied for any reasonable sample size.

We expect also that the results of the dc transformer measurements in thick
samples
($d \gtrsim \Lambda$) should be qualitatively different from the samples
reported in \cite{safar,bariloche} due to the screening effects.
Finally, we note that the results obtained in twinned samples for the tilted
field will be severely modified by the interaction between vortices
and twins because in these conditions each vortex line intersects the twin
boundary which affects strongly the motion and equilibrium positions of vortex
lines.

In conclusion, the observations \cite{safar,bariloche} strongly support the
existence of a new vortex line liquid state sandwiched between the
normal state and the vortex glass.

We are grateful to H. Safar for useful discussions. We acknowledge the support
of the innovation partnership grant of the State of NJ and  of the grant M6M000
from International Science Foundation.

\end{multicols}

\end{document}